\documentclass[onecolumn]{article}
\usepackage{graphicx}
\usepackage[utf8]{inputenc}
\usepackage{amsmath}
\usepackage{physics}
\usepackage{mathtools}
\usepackage{amssymb}
\usepackage{slashed}
\usepackage{tikz-feynman}
\usepackage{amsthm}
\usepackage{txfonts}
\usepackage{geometry}
\begin{document}

\title{Why Neutrino Masses Cannot Arise from Nonzero VEV of Charged Higgs Field in the Only Higgs Doublet}
\author{Lu Jianlong\\ \emph{\small{Department of Physics, National University of Singapore}}}
\date{}
\maketitle

\begin{abstract}
A scheme of neutrino mass generation is proposed by \cite{wrong}, in which the small nonzero neutrino masses come from nonzero vacuum expectation value (VEV) of the charged Higgs field in the Higgs doublet in the Standard Model of particle physics (SM). However, the introduction of nonzero VEV for the charged Higgs field implies broken ${\rm U}(1)$ symmetry in the electroweak theory, which leads to non-conservation of electric charge and nonzero mass of photon. By comparing the predicted ratio of VEVs of charged Higgs field and neutral Higgs field in \cite{wrong} with the ratio constrained by the present experimental upper bound of photon mass, we show that this scheme does not work. 
\end{abstract}

\section{INTRODUCTION}
Neutrinos (antineutrinos) are assumed to be massless in the unextended SM, with only left-handed (right-handed) components appearing in the SM Lagrangian density. Nonetheless, flavor-changing phenomena of neutrinos have been observed since Homestake experiment in late 1960s. \cite{Homestake} One natural solution is to introduce nonzero neutrino masses such that there is a mismatch between neutrino mass eigenstates and neutrino flavor eigenstates. In the SM, all masses of massive elementary particles, including fermions and gauge bosons of weak interaction, arise from Higgs mechanism with one Higgs doublet consisting of one neutral Higgs field and one charged Higgs field. With a nonzero VEV in the neutral Higgs field, the gauge symmetry group ${\rm SU}(2)_{L}\times {\rm U}(1)_{Y}$ is broken to ${\rm U}(1)_{Q}$ so that three gauge bosons in weak interaction gain nonzero masses. The zero VEV of charged Higgs field ensures the conservation of electric charge and that the remaining gauge boson for electromagnetic interaction, i.e., photon, remains massless.\\
In \cite{wrong}, the authors argue that neutrino masses can be generated by introducing nonzero VEV for the charged Higgs field, which are uniquely related to the masses of charged leptons with the VEVs of Higgs doublet as the only parameters. In the following analysis, we show that this scheme does not work, since it predicts a ratio of VEVs of charged Higgs field and neutral Higgs field much larger than the ratio constrained by the present experimental upper bound of photon mass.

\section{CALCULATION}
In the unextended SM, only the neutral component of the Higgs doublet has nonzero VEV, i.e.,
\begin{equation}
    \Phi_{VEV}=\frac{1}{\sqrt{2}}\begin{pmatrix} 0\\ v\end{pmatrix}
\end{equation}
with $v\neq 0$. Now we suppose the charged Higgs field also has nonzero VEV, as proposed in \cite{wrong},
\begin{equation}
    \Phi_{VEV}=\frac{1}{\sqrt{2}}\begin{pmatrix} v^{+}\\ v\end{pmatrix}
\end{equation}
with $v^{+},v\neq 0$.\\
In the unitary gauge, the Higgs doublet is as follows, with only one physical scalar boson surviving,
\begin{equation}
   \Phi=\frac{1}{\sqrt{2}}\begin{pmatrix}v^{+}\\ v+h(x)\end{pmatrix}.
\end{equation}
The kinetic term $(D^{\mu}\Phi)^{\dagger}D_{\mu}\Phi$ of Higgs doublet is defined by the covariant derivative minimally coupling with four vector fields $(W^{1}_{\mu},W^{2}_{\mu},W^{3}_{\mu},B_{\mu})$, \cite{Higgs}\cite{XZZ}
\begin{equation}
   D_{\mu}\Phi=\Big(\partial_{\mu}-\sum_{a=1}^{3}ig_{2}\frac{\sigma_{a}}{2}W^{a}_{\mu}-ig_{1}\frac{Y}{2}B_{\mu}\Big)\Phi
\end{equation}
in which $g_{2}$ and $g_{1}$ are coupling constants of ${\rm SU}(2)$ and ${\rm U}(1)$ respectively, $\sigma_{a}$ are Pauli matrices while $Y$ is the identity matrix. Recall that Pauli matrices are 
\begin{equation}
    \sigma_{1}=\begin{pmatrix} 0 & 1\\ 1 & 0\end{pmatrix},\ \ \ \sigma_{2}=\begin{pmatrix} 0 & -i \\ i & 0\end{pmatrix},\ \ \ \sigma_{3}=\begin{pmatrix} 1 & 0\\ 0 & -1\end{pmatrix}.
\end{equation}
By making use of the explicit forms of Pauli matrices and $\Phi$, we obtain the following expression of the kinetic term,
\begin{equation}
\resizebox{.9 \textwidth}{!} 
{
$\begin{gathered}
    (D^{\mu}\Phi)^{\dagger}D_{\mu}\Phi=\Big[\Big(\partial_{\mu}-\sum_{a=1}^{3}ig_{2}\frac{\sigma_{a}}{2}W^{a}_{\mu}-ig_{1}\frac{Y}{2}B_{\mu}\Big)\Phi\Big]^{\dagger} \Big(\partial_{\mu}-\sum_{a=1}^{3}ig_{2}\frac{\sigma_{a}}{2}W^{a}_{\mu}-ig_{1}\frac{Y}{2}B_{\mu}\Big)\Phi\\
   =\Big( \partial^{\mu}\Phi^{\dagger}+ \sum_{a=1}^{3}ig_{2}\Phi^{\dagger}\frac{\sigma_{a}^{\dagger}}{2}W_{a}^{\mu}+ ig_{1}\Phi^{\dagger}\frac{Y^{\dagger}}{2}B^{\mu}\Big) \Big(\partial_{\mu}\Phi-\sum_{a=1}^{3}ig_{2}\frac{\sigma_{a}\Phi}{2}W^{a}_{\mu}-ig_{1}\frac{Y\Phi}{2}B_{\mu}\Big)\\
   = \frac{1}{\sqrt{2}}\begin{pmatrix} i\frac{g_{2}(v^{*}+h^{*})W_{1}^{\mu}}{2}- \frac{g_{2}(v^{*}+h^{*})W_{2}^{\mu}}{2}+ i\frac{g_{2}v^{+*}W_{3}^{\mu}}{2} + i\frac{g_{1}v^{+*}B^{\mu}}{2}     & \partial^{\mu}h^{*}+ i\frac{g_{2}v^{+*}W_{1}^{\mu}}{2}+ \frac{g_{2}v^{+*}W_{2}^{\mu}}{2}- i\frac{g_{2}(v^{*}+h^{*})W_{3}^{\mu}}{2}+ i\frac{g_{1}(v^{*}+h^{*})B^{\mu}}{2} \end{pmatrix} \times\frac{1}{\sqrt{2}}\begin{pmatrix} -i\frac{g_{2}(v+h)W_{\mu}^{1}}{2}- \frac{g_{2}(v+h)W_{\mu}^{2}}{2} -i\frac{g_{2}v^{+}W_{\mu}^{3}}{2} -i\frac{g_{1}v^{+}B_{\mu}}{2}  \\ \partial_{\mu}h-  i\frac{g_{2}v^{+}W_{\mu}^{1}}{2} + \frac{g_{2}v^{+}W_{\mu}^{2}}{2}+ i\frac{g_{2}(v+h)W_{\mu}^{3}}{2} - i\frac{g_{1}(v+h)B_{\mu}}{2} \end{pmatrix}\\
   = \frac{1}{\sqrt{2}} \begin{pmatrix} i\frac{g_{2}(v^{*}+h^{*})(W_{1}^{\mu}+iW_{2}^{\mu})}{2}+ i\frac{v^{+*}(g_{2}W_{3}^{\mu}+g_{1}B^{\mu})}{2}      & \partial^{\mu}h^{*}+ i\frac{g_{2}v^{+*}(W_{1}^{\mu}-iW^{\mu}_{2})}{2}- i\frac{(v^{*}+h^{*})(g_{2}W_{3}^{\mu}-g_{1}B^{\mu})}{2} \end{pmatrix} \times\frac{1}{\sqrt{2}}\begin{pmatrix} -i\frac{g_{2}(v+h)(W_{\mu}^{1}-iW^{2}_{\mu})}{2} -i\frac{v^{+}(g_{2}W_{\mu}^{3}+g_{1}B_{\mu})}{2}   \\ \partial_{\mu}h-  i\frac{g_{2}v^{+}(W_{\mu}^{1}+iW^{2}_{\mu})}{2} + i\frac{(v+h)(g_{2}W_{\mu}^{3}-g_{1}B_{\mu})}{2}  \end{pmatrix}\\
   = \frac{g_{2}^{2}|v+h|^{2}}{8} (W_{1}^{\mu}+iW_{2}^{\mu}) (W_{\mu}^{1}-iW^{2}_{\mu})+ \frac{|v^{+}|^{2}}{8} (g_{2}W^{\mu}_{3}+g_{1}B^{\mu}) (g_{2}W_{\mu}^{3}+g_{1}B_{\mu})
    + \frac{1}{2}( \partial^{\mu}h^{*})   \partial_{\mu}h-  i ( \partial^{\mu}h^{*})  \frac{g_{2}v^{+}(W_{\mu}^{1}+iW^{2}_{\mu})}{4} + i ( \partial^{\mu}h^{*})  \frac{(v+h)(g_{2}W_{\mu}^{3}-g_{1}B_{\mu})}{4}\\
    + i(\partial_{\mu}h)\frac{g_{2}v^{+*}(W_{1}^{\mu}-iW^{\mu}_{2})}{4}+ \frac{g_{2}^{2}|v^{+}|^{2}}{8}(W^{\mu}_{1}-iW^{\mu}_{2})(W_{\mu}^{1}+iW^{2}_{\mu})- i (\partial_{\mu}h)\frac{(v^{*}+h^{*})(g_{2}W_{3}^{\mu}-g_{1}B^{\mu})}{4}+ \frac{|v+h|^{2}}{8}(g_{2}W^{\mu}_{3}-g_{1}B^{\mu} )(g_{2}W_{\mu}^{3}-g_{1}B_{\mu}).
\end{gathered}$
}
\end{equation}
The four gauge boson fields in electroweak interaction, including two charged ones ($W^{+}_{\mu}$ and $W^{-}_{\mu}$) and two neutral ones ($Z_{\mu}$ and $A_{\mu}$), are defined as 
\begin{equation}
   W^{\pm}_{\mu}=\frac{1}{\sqrt{2}}(W_{\mu}^{1}\mp iW_{\mu}^{2}),
\end{equation}
\begin{equation}
    Z_{\mu}=\frac{g_{2}W_{\mu}^{3}-g_{1}B_{\mu}}{\sqrt{g_{2}^{2}+g_{1}^{2}}},
\end{equation}
\begin{equation}
    A_{\mu}=\frac{g_{2}W_{\mu}^{3}+g_{1}B_{\mu}}{\sqrt{g_{2}^{2}+g_{1}^{2}}}.
\end{equation}
We choose positive real $v$ and real $h$ for convenience. The mass terms of four types of gauge bosons in the above kinetic term are 
\begin{equation}
    \frac{g_{2}^{2}(v^{2}+|v^{+}|^{2})}{8} (W_{1}^{\mu}+iW_{2}^{\mu}) (W_{\mu}^{1}-iW^{2}_{\mu})+ \frac{|v^{+}|^{2}}{8} (g_{2}W^{\mu}_{3}+g_{1}B^{\mu}) (g_{2}W_{\mu}^{3}+g_{1}B_{\mu})+ \frac{v^{2}}{8}(g_{2}W^{\mu}_{3}-g_{1}B^{\mu} )(g_{2}W_{\mu}^{3}-g_{1}B_{\mu}).
\end{equation}
Their masses can be directly read off, which are 
\begin{equation}
    m_{W^{+}}=m_{W^{-}}=\frac{g_{2}\sqrt{v^{2}+|v^{+}|^{2}}}{2},\ \ \ \ m_{Z}=\frac{v\sqrt{g_{2}^{2}+g_{1}^{2}}}{2},\ \ \ \ m_{A}=\frac{|v^{+}|\sqrt{g_{2}^{2}+g_{1}^{2}}}{2}.
\end{equation}
When $v^{+}=0$, they reduce to the original results in the SM,
\begin{equation}
    m_{W^{+}}=m_{W^{-}}=\frac{g_{2}v}{2},\ \ \ \ m_{Z}=\frac{v\sqrt{g_{2}^{2}+g_{1}^{2}}}{2},\ \ \ \ m_{A}=0.
\end{equation}
In the double VEV scheme we discuss above, the ratio of masses between photon and $Z$ boson is 
\begin{equation}
    \frac{m_{A}}{m_{Z}}=\frac{|v^{+}|}{v}.
\end{equation}
The best estimation of photon mass at the present stage is of the order of magnitude $<10^{-18}eV/c^{2}$, while the mass of $Z$ boson is about $(91.1876\pm 0.0021 )\times 10^{9}eV/c^{2}$. \cite{pm1} Hence, the ratio of masses between photon and $Z$ boson (and thus the ratio between $|v^{+}|$ and $v$) is at least constrained by 
\begin{equation}
     \frac{|v^{+}|}{v}= \frac{m_{A}}{m_{Z}}< 10^{-27}.
\end{equation}
For comparison, the authors in \cite{wrong} have the following estimation about ratio between $|v^{+}|^{2}$ and $v^{2}$ based on experimental data on charged lepton masses and neutrino mass-squared differences,
\begin{equation}
    \frac{|v^{+}|^{2} }{v^{2}}\sim 10^{-11}.
\end{equation}
The huge discrepancy obviously indicates that the scheme proposed in \cite{wrong} does not work.

\section{CONCLUSION}
We have calculated the mass of photon predicted by Higgs mechanism with an additional nonzero VEV $v^{+}$ of charged Higgs field in the only Higgs doublet. The present experimental data on photon mass constrains the ratio between $|v^{+}|$ and $v$ to be less that $10^{-27}$, which is much smaller than the result given in \cite{wrong} based on charged lepton masses and neutrino mass-squared differences. Therefore, we can conclude that the neutrino masses should not come from nonzero VEV of charged Higgs field in the only Higgs doublet.


\end{document}